\documentclass[12pt]{article}
\usepackage{epsf,amsfonts}
\def\epsfsize#1#2{1.0#1}
\epsfverbosetrue
\def\goth{\mathfrak}          
\def\double{\mathbb}         
\def\ccal{\cal}           

\def\cc{{\double C}}     
       
\def\rr{{\double R}}

\def\aa{{\cal A}}
\def\ccc{{\cal C}}
\def\dd{{\cal D}} 
\def\gg{{\goth g}}        
\def\hh{{\cal H}}
\def\hhh{{{\double H}}}   
\def\ff{{\cal F}}
\def\mm{{{\ccal M}}}

\def\aa{{\cal A}}
\def\dd{{\cal D}} 
\def\hh{{\cal H}}
\def\ff{{\cal F}}
\def\lll{{\cal L}}
\def\sss{{\cal S}}
\def\jj{{\cal J}}
\def\t{{\rm tr}\,} 
\def\re{{\rm Re}\,}

\def\ddd{{\,\hbox{$\partial\!\!\!/$}}}
\def\ddee{{\,\hbox{${\rm D}\!\!\!\!/\,$}}} 
\def\aaa{{\,\hbox{$A\!\!\!/\,$}}} 
\def\dee{\hbox{\rm D}}
\def\de{\hbox{\rm d}}

\def\ot{\otimes}
\def\op{\oplus}

\def\bb{\begin{eqnarray}}
\def\ee{\end{eqnarray}}
\def\eee{\nonumber\end{eqnarray}}
\def\pp{\pmatrix}
\def\qq{\quad}

\begin{document}

\hsize 17truecm
\vsize 24truecm
\font\twelve=cmbx10 at 13pt
\font\eightrm=cmr8
\baselineskip 18pt

\begin{titlepage}

\centerline{\twelve CENTRE DE PHYSIQUE TH\'EORIQUE}
\centerline{\twelve CNRS - Luminy, Case 907}
\centerline{\twelve 13288 Marseille Cedex 9}
\vskip 3truecm

\centerline{\twelve NONCOMMUTATIVE YANG-MILLS AND }
\centerline{\twelve NONCOMMUTATIVE RELATIVITY: }
\centerline{\twelve A BRIDGE OVER TROUBLE WATER}

\bigskip

\begin{center}
{\bf Lionel CARMINATI}
\footnote{\, and Universit\'e de Provence\\
carminati@cpt.univ-mrs.fr \qq\qq 
\qq iochum@cpt.univ-mrs.fr \qq\qq
\qq schucker@cpt.univ-mrs.fr} \\
\bf Bruno IOCHUM$^{\,1}$\\
\bf Thomas SCH\"UCKER $^{1}$
\end{center}

\vskip 2truecm
\leftskip=1cm
\rightskip=1cm
\centerline{\bf Abstract} 

\medskip

Connes' view at Yang-Mills theories is reviewed with
special emphasis on the gauge invariant scalar
product. This landscape is shown to contain
Chamseddine and Connes' noncommutative extension
of general relativity restricted to flat space-time, if
the top mass is between 172 and 204 GeV. Then the
Higgs mass is between 188 and 201 GeV.

\vskip 1truecm
PACS-92: 11.15 Gauge field theories\\ 
\indent
MSC-91: 81T13 Yang-Mills and other gauge theories 
 
\vskip 1truecm

\noindent june 1997
\vskip 1truecm
\noindent CPT-96/P.3477\\
\noindent hep-th/yymmxxx
 
\vskip1truecm

 \end{titlepage}

\section{Introduction}
 
 Einstein was a passionate sailor. We speculate that
this was no accident. The subtle harmony between
geometries and forces becomes palpable to the sailor,
he sees the curvature of the sail and feels the force
that it produces. Before Einstein, it was generally
admitted that forces are vector fields in an Euclidean
space, $\rr^3$, the scalar product being necessary to
define work and energy. Einstein generalized
Euclidean to Minkowskian and Riemannian geometry
and we have two {\it dreis\"atze} or {\it r\`egles de
trois}. Take Coulomb's static law for the electric field
with coupling constant
$\epsilon_0$ and add Minkowskian geometry with its
scale
$c$, the speed of light: you obtain Maxwell's
theory. In particular, there appears the magnetic
field with feeble coupling constant
$\mu_0=1/(c^2\epsilon_0)$. Maxwell's theory is
celebrated today as Abelian or should we say,
commutative Yang-Mills theory. The second {\it
dreisatz} starts from Newton's (static) universal law of
gravitation, adds Riemannian geometry to obtain
general relativity with new feeble, gravito-magnetic
forces.

Connes proposes two more dreis\"atze.
Take a certain Yang-Mills theory with coupling
constant $g$, coupled to a Dirac spinor of mass $m$.
Add noncommutative geometry \cite{book} with an
energy scale $\Lambda$: you obtain a
Yang-Mills-Higgs theory
\cite{cl},\cite{tresch}. The symmetry breaking scalar
becomes a magnetic field of the Yang-Mills field and
its mass and self-coupling $\lambda$ are constrained
in terms of
$g$, $m$ and $\Lambda$. His second dreisatz starts
from general relativity, adds noncommutative
geometry to obtain the Einstein-Hilbert action plus
the Yang-Mills-Higgs action \cite{grav}\cite{cc}. Now
the Yang-Mills and the Higgs fields are magnetic
fields of the gravitational field. Again there are
constraints on $\lambda$, but they are different.  

Let us call noncommutative Yang-Mills the third
and noncommutative relativity the fourth
dreisatz. Note however that --- unlike with
supersymmetry --- you cannot take any Yang-Mills
theory and put 'noncommutative' in front
\cite{versus}\cite{tk}\cite{mz}. Note also that, behind 
noncommutative relativity, there stands a
genuine noncommutative extension of Einstein's
principle of general relativity, the spectral principle.

One of the attractive features of noncommutative
geometry is to unify gauge couplings with scalar
self-couplings and Yukawa couplings. These 
couplings will be at the center of our discussion. They
are related to the set of all scalar products on a given
space,which is a cone: in the case of gauge invariant
scalar products on the Lie algebra of a Yang-Mills
theory, the gauge couplings are positive coordinates
on this cone. For us, a noncommutative
geometry is given by a spectral triple $(\aa,\hh,\dd)$.
The motivation for this noncommutative geometry
comes from Connes' theorem \cite {grav} establishing
a one-to-one correspondence between {\it
commutative} spectral triples and Riemannian spin
geometries \cite{other}.
$\aa$ is an associative involution algebra with unit,
represented faithfully on a Hilbert space $\hh$ on
which also the self-adjoint 'Dirac' operator $\dd$ acts.
Physically, the Hilbert space is spanned by the
fermions. The Dirac operator serves several purposes:
it defines the kinetic term for fermions, it allows to
construct differential forms --- the Yang-Mills fields
are 1-forms ---, it is used as ultra-violet regulator to
define the scalar product and above all, it defines the
metric structure of space-time.

So, in this paper, playing with the scalar products and the experimental accuracies of 
the gauge couplings and the masses, we consider the noncommutative Yang-Mills and 
noncommutative relativity theories as effective theories, using the renormalization flow 
and show that they can be related according to certain constraints on the Higgs and top 
masses.

\section{Noncommutative relativity}

The dynamical variable of gravity is the metric on
space-time. Einstein used the matrix
$g^{\mu\nu}(x)$ of the metric $g$ with respect to a
coordinate system $x^\mu$ to parameterize the set of
all metrics on a fixed space-time $M$. The coordinate
system being unphysical, Einstein required his field
equations for the metric to be covariant under 
coordinate transformations, the principle of general
relativity. Elie Cartan used tetrads, {\it rep\`eres
mobiles}, to parameterize the set of all metrics. This
parameterization allows to generalize the Dirac
operator $\dd$ to curved space-times and also
reformulates general relativity as a gauge theory
under the Lorentz group. Connes
\cite{grav} goes one step further by relating the set
of all metrics to the set of all Dirac operators. The
Einstein-Hilbert action, from this point of view, is the
Wodzicki residue of the second inverse power of the
Dirac operator
\cite{wod} and is computed most conveniently from
the second coefficient of the heat kernel expansion of
the Dirac operator squared. The heat kernel expansion
\cite{heat}
is an old friend \cite{friend} from quantum field
theory in curved space-time, from its formal relation to
the one-loop effective action 
\bb S_{\it eff}=\t \log (\dd^2/\Lambda^2),\qquad
\Lambda\ {\rm a\ cutoff}.\ee 
This relation has been used by
Sakharov \cite{sak} to induce gravity from quantum
fluctuations, leading however to a negative Newton
constant \cite{sakcor}. 

By generalizing the metric, the Dirac operator plays a
fundamental role in noncommutative geometry. To
describe Yang-Mills theories, Connes considers
the tensor product of space-time and internal
space in this new geometry, a natural point of view
because the fermionic mass matrix qualifies as Dirac
operator on internal space. This cheap tensor product
unifies space-time diffeomorphisms with internal gauge
transformations by extending
Einstein's principle of general relativity to 
noncommutative geometry. Remember that Einstein
constructed general relativity in two steps, by
applying his principle first to matter, then to the
gravitational field itself. Connes follows this pattern
and of course in his case, the spinors are the matter.
 
 To generalize the Dirac
operator from flat to curved space-time (locally), it is 
sufficient to write the Dirac operator first in flat
space-time but with respect to noninertial coordinates.
A straightforward calculation produces the {\it
covariant} Dirac operator that contains the spin
connection $\omega$. Although of vanishing
curvature, $\omega$ contains already a lot of physics,
e.g. the centrifugal and Coriolis accelerations in the
coordinates of the rotating disk, the quantum
interference pattern of neutrons \cite{bonse} in
oscillating coordinates. Then, the generalization to
curved space is easy where $\omega$ describes the
(minimal) coupling of the spinor to the gravitational
field. In Einstein's spirit, the covariant Dirac
operator is obtained by acting with the
diffeomorphism group on the flat Dirac operator. But
the diffeomorphism group is just the automorphism
group of the associative (and commutative) algebra
${\cal C}^\infty(M)$ representing space-time in
noncommutative geometry. On the other hand, the
product of space-time and internal space is represented
in this geometry by the tensor product of ${\cal
C}^\infty(M)$ with a matrix algebra. Its automorphism
group is the semi-direct product of the
diffeomorphisms and unitaries of the matrix algebra,
the internal gauge transformations. The
diffeomorphisms are the outer, the gauge
transformations are the inner automorphisms. And
what do we get when this entire automorphism group
acts on the flat Dirac operator? We get the total
covariant Dirac operator containing the spin
connection, the gauge connection and the Higgs
\cite{tresch}. In other words, we get the minimal
couplings of the Dirac spinor to the gravitational and
Yang-Mills fields and its Yukawa couplings to the
Higgs field. In Connes' words, the Higgs and
Yang-Mills fields are noncommutative fluctuations of
the metric. (Abelian Yang-Mills theories do not have
such fluctuations.) Accordingly, Connes generalizes
Einstein's principle of general relativity by
postulating that only the intrinsic properties of the
covariant Dirac operator be relevant for physics. Here
intrinsic means invariant under automorphisms.
Thus, these properties must concern the spectrum
only. Spectral principle is a convenient name for
Connes' generalization of the principle of general
relativity.

So far, we have only the kinematic of the metric (and
its fluctuations). To get its dynamics, Einstein
developed the full power of the principle of general
relativity and derived the Einstein-Hilbert action. In
short this story: the $1/r^2$ in Newton's universal law 
is the Green function of the divergence, an operator of
first order in the forces. We already know that the
forces are encoded in the connection $\omega$.
Riemannian geometry tells us that the connection is
obtained from first order derivatives of the metric.
Therefore Einstein looked for a second order
differential equation for the metric. The covariance
under change of coordinates fixes this equation up to
the cosmological constant to be the Einstein equation.
 Chamseddine \& Connes \cite{cc} reedit this story
using the spectral principle. It is stronger than
Einstein's principle in the sense that for the metric
only, the Einstein-Hilbert action follows without the
use of Newton's law. In addition, the spectral principle
fixes the action of the fluctuations to be the
Yang-Mills action, the covariant Klein-Gordon
action and the symmetry breaking Higgs potential. 
Warning: following physicists' habits, we have 
confused diffeomorphisms and coordinate
transformations. Cleaning up this point leads to deep
mathematics \cite{cm} and probably a further
unification of general relativity and Yang-Mills
theory: the reduction of the diffeomorphism group to
an isometry group might take the form of a
spontaneous symmetry break down.

\subsection{The stiff action}

In even dimensions, the spectrum of the Dirac
operator is even and it is sufficient to
consider the positive part of the spectrum which in
the  Euclidean is conveniently characterized by a
distribution function
\bb S=\t f(\dd^2/\Lambda^2),\ee
where $\Lambda$ is an energy cutoff and 
$f:\rr_+\rightarrow\rr_+$ is a positive, smooth
function with finite, strictly positive first `momenta',
\bb f_0:=\int_0^\infty uf(u)\,\de u,\qq
f_2:=\int_0^\infty f(u)\,\de u,\qq
f_4:=f(0).\ee
If instead,
$f$ was the logarithm, this trace, after a proper
renormalization, would be Sakharov's induced gravity
action.
The positive function $f$ is universal: the action $S$
can be computed asymptotically \cite{egbv},
that is up to terms of the order of $\Lambda^{-2}$,
using the Lichn\'erowicz formula and the heat kernel
expansion. The action depends only on the three
momenta
$f_0,\ f_2,\ f_4$ and takes the form:
\bb \t f(\dd_{t,{\rm cov}}^2/\Lambda^2)&\approx&\int_M[
-{\textstyle\frac{1}{16\pi}}m_P(\Lambda)^{2}\,R
\,+\,\Lambda_C(\Lambda)\cr &&\qq+\,
{\textstyle\frac{1}{2}}g_3(\Lambda)^{-2}\, \t
F_{\mu\nu}^{(3)}F^{(3)\mu\nu}+
{\textstyle\frac{1}{2}}g_2(\Lambda)^{-2}\, \t
F_{\mu\nu}^{(2)}F^{(2)\mu\nu}+
{\textstyle\frac{1}{4}}g_1(\Lambda)^{-2}\, 
F_{\mu\nu}^{(1)}F^{(1)\mu\nu}\cr &&\qq
+\,{\textstyle\frac{1}{2}}\,(\dee_\mu\varphi)^*
\dee^\mu\varphi\,+\,\lambda(\Lambda) |\varphi|^4
\,-\,{\textstyle\frac{1}{2}}\,\mu(\Lambda)^2|\varphi|^2
\cr &&\qq-\,a(\Lambda)
C_{\mu\nu\rho\sigma}C^{\mu\nu\rho\sigma}\,+\,
{\textstyle\frac{1}{12}}\,|\varphi|^2
R\qq
]\,(\det
g_{\cdot\cdot})^{1/2}\,\de^4x.\label{action}\ee  
Here
$\dd_{t,{\rm cov}}$ is the total, covariant Dirac
operator of the standard model of electroweak and
strong interactions with
$N=3$ generations of quarks and leptons. It follows
that $\varphi$ is an isospin doublet. After a proper
normalization of the kinetic terms and a shift of the
Higgs field by its vacuum expectation value,
$|\varphi|=v(\Lambda)=\mu(\Lambda)/(2\sqrt\lambda(\Lambda))$, we can
identify Newton's constant
$G=16\pi\hbar c\,m_P^2$, the cosmological
constant
$\Lambda_C$ and the other coupling constants 
\bb
m_P(\Lambda)^2&=&{\textstyle\frac{1}{3\pi}}f_2\left[15N
 -2\frac{L^2}{Q}\,\right]
\Lambda^2,
\\  &&L(\Lambda)\
:=\ 3(m^2_t+m^2_c+m^2_u+m^2_b+m^2_s+m^2_d)+
m^2_\tau+m^2_\mu+m^2_e,\\
&&Q(\Lambda)\ :=\ 
3(m^4_t+m^4_c+m^4_u+ m^4_b+m^4_s+m^4_d)+
m^4_\tau+m^4_\mu+m^4_e,\\
\Lambda_C(\Lambda)&=&{\textstyle\frac{1}{4\pi^2}}
\left[15Nf_0
-\,\frac{f_2^2}{f_4}\,\frac{L^2}{Q}\,\right]
\Lambda^4,\\
g_3(\Lambda)^{-2}&=&{\textstyle\frac{N}{3\pi^2}}f_4,
\label{stiff3}\\ 
g_2(\Lambda)^{-2}&=&{\textstyle\frac{N}{3\pi^2}}f_4,\\ 
g_1(\Lambda)^{-2}&=&{\textstyle\frac{5}{3}}
{\textstyle\frac{N}{3\pi^2}}f_4,\label{stiff1}\\ 
\lambda(\Lambda)^{-1}&=&{\textstyle\frac{1}{\pi^2}}\;f_4\;\frac{L(\Lambda)^2}
{Q(\Lambda)}\,=\,{\textstyle\frac{3}{\pi^2}}\; f_4
(1+2\frac{m_b(\Lambda)^2}{m_t(\Lambda)^2}\,+
\,O(\frac{m_{\tau}(\Lambda)^2}{m_t(\Lambda)^2})),\\
\mu(\Lambda)^2&=& 2\,\frac{f_2}{f_4}\,\Lambda^2
\label{stiffmu},\\
a(\Lambda)&=&{\textstyle\frac{3N}{64\pi^2}}\ f_4.\ee
From now on, we ignore the
gravitational part because we want to use the
renormalization flow of the coupling constants and
also because we want to compare this theory with
the noncommutative Yang-Mills that, by the way,
automatically has a vanishing cosmological constant
as we shall see. 

The constraints for the gauge couplings,
\bb
g_3(\Lambda)=g_2(\Lambda) \ {\rm and}\ 
\sin^2\theta_w(\Lambda)=\frac{g_2^{-2}}{g_1^{-2}+g_2^{-2}}=\frac{3}{8}
\eee 
(\ref{stiff3}-\ref{stiff1}), are familiar from
grand unification and force us to assume the big
desert. Consequently all numerical considerations will
be qualitative only. Indeed, the three gauge couplings
$g_i(\Lambda)$, once fixed at the $Z$-mass to their
experimental values $\sin^2\theta_w(m_Z)=0.2315\,\pm\,0.0005$, 
see appendix, 
do not meet in a point anymore as was the case in
the
$SU(5)$ days. Today they define a triangle with
 $\Lambda= 10^{13}-10^{17}$ GeV and
$\sqrt{\frac{5}{3}}g_1(\Lambda), g_2(\Lambda), g_3(\Lambda)$ are in the interval $0.52-0.56$, 
Figure \ref{fig:The coupling constants}. Details on the
renormalization group flow can be found in the
appendix. For the noncommutative constraints
(\ref{stiff3}-\ref{stiff1}), this means that $f_4$ cannot
take a precise value, ${\textstyle\frac{1}{4\pi^2}}\;f_4=0.80-0.94$.

\subsection{The soft Einstein-Hilbert action}

Of course, we may try to do better by introducing more
parameters. Let
$z'$, the `noncommutative coupling constant', be a
positive operator on the fermionic Hilbert space that
commutes with the representation and the Dirac
operator. For the standard model, this $z'$ contains
four positive numbers $x'$, $y'_1$, $y'_2$,
$y'_N$. We soften the action (\ref{action}) to $ \t[z'
f(\dd_{t,cov}^2/\Lambda^2)]$. Then the constraints
read
\cite{ciks}:
\bb
g_3(\Lambda)^{-2}&=&{\textstyle\frac{1}{9\pi^2}}f_4Nx',
\label{soft3}\\ 
g_2(\Lambda)^{-2}&=&{\textstyle\frac{1}{12\pi^2}}f_4(Nx'+
y'_1+y'_2+y'_N),\\ 
g_1(\Lambda)^{-2}&=&{\textstyle\frac{1}{12\pi^2}}f_4(
{\textstyle\frac{11}{9}}Nx'+3(y'_1+y'_2+y'_N))
\label{soft1},\\ 
\lambda(\Lambda)^{-1}&=&{\textstyle\frac{1}{\pi^2}}
f_4\frac{L(\Lambda)^2}{Q(\Lambda)}
\,\label{softlam}
\\  &&L(\Lambda)\  =\ 
x'(m^2_t+m^2_c+m^2_u+m^2_b+m^2_s+m^2_d)+
y'_3m^2_\tau+y'_2m^2_\mu+y'_1m^2_e,\\
&&Q(\Lambda)\ =\  x'(m^4_t+m^4_c+m^4_u+
m^4_b+m^4_s+m^4_d)+
y'_3m^4_\tau+y'_2m^4_\mu+y'_1m^4_e,\\
\mu(\Lambda)^2&=& 2\,\frac{f_2}{f_4}\,\Lambda^2
\label{softmu}\ee
If $z'=1_{90}$, then $x'=3,\ y'_1=y'_2=y'_N=1$ and we
recover the stiff relations (\ref{stiff1}-\ref{stiffmu}).

\subsection{The dominating top approximation and
renormalization flow}

The soft relations do not have the problem
$g_3(\Lambda)=g_2(\Lambda)$ anymore, but we still
cannot avoid the desert. In fact now 
\bb 
sin^2\theta_w(\Lambda)\ =\
\frac{Nx'+(y'_1+y'_2+y'_N)}
{{\textstyle\frac{20}{9}}Nx'+4(y'_1+y'_2+y'_N)},
\eee
and the weak mixing angle is constrained for all $\Lambda$: ${\textstyle\frac{1}{4}}\,<\,
\sin^2\theta _w(\Lambda)\,<\,{\textstyle\frac{9}{20}}$.
From now on, we neglect all fermion masses with
respect to the top mass. This approximation induces
relative errors of the order of $m_b^2/m_t^2=0.0006$
and it reduces the number of positive parameters from
four, $x',\ y'_1,\ y'_2,\ y'_N$ to two, $x'$ and $y':=
y'_1+y'_2+y'_N$. In the one loop
approximation, the evolution of the gauge couplings
(\ref{g1evolution}-\ref{g3evolution}) decouples from the other couplings and we can solve
the constraints (\ref{soft3}-\ref{soft1}) such that at
the $Z$ mass, they reproduce precisely the
experimental values. The last
non-empty constraint (\ref{softlam}),
$\lambda(\Lambda)=
{\textstyle\frac{N}{9}}g_3(\Lambda)^2$ then fixes the
Higgs mass. With these
approximations, we obtain:

- In the stiff case, $x'=y'=3$, the incertainty on the cutoff is large:
\[
\left.
\begin{array}{ll}
\Lambda=(10^{13}-10^{17})\ {\rm GeV}, \\ \\
{\textstyle\frac{1}{12\pi^2}}f_4x'=
{\textstyle\frac{1}{12\pi^2}}f_4y'=0.80-0.94,\\ \\
m_H= 182\pm 10\pm 2\pm 7\ {\rm GeV}.
\end{array}
\right\}{\rm stiff \  EH}
\]

The first error is from the uncertainty in the
noncommutative scale $\Lambda$, the second from
the present experimental uncertainty in the gauge 
couplings,
$g_3=1.218\pm 0.026$, and the third from the
uncertainty in the top mass,
$m_t=175\pm 6$ GeV.

- In the soft case, the cutoff is
sharp:
\[
\left.
\begin{array}{ll}
\Lambda=0.96\cdot 10^{10}\ {\rm GeV}, \\ \\
{\textstyle\frac{1}{12\pi^2}}f_4x'=0.578,\\  \\
{\textstyle\frac{1}{12\pi^2}}f_4y'= 1.369\\ \\
m_H=190\pm 0\pm 1\pm 4\ {\rm GeV}.
\end{array}
\right\}{\rm soft \  EH}
\]
Let us anticipate
that this comparison will work quantitatively only for
the stiff case. So if we spell out the soft case here, then not because we
believe that it makes sense to fit numbers through the
big desert with the indicated precision. Our aim is to
assess the stability of the Higgs mass prediction and
also to make the comparison with the
noncommutative Yang-Mills easier.

\section{ The noncommutative Yang-Mills action}

After this quick review of the noncommutative
version of general relativity in flat space-time, we now
turn to our main concern,
noncommutative Yang-Mills theory, Connes' first
dreisatz. The point will be that Connes' second dreisatz
adds insight to the older one.

\subsection{The conventional scalar products}

To construct a Yang-Mills action $\int\t F*F$, we need
four ingredients: differential forms on space-time $M$,
a Lie group $G$, `the internal space', a scalar product
on the space of differential forms $\Omega M$ and an
invariant scalar product on the Lie algebra $\gg$ of
the group $G$. The gauge field $A$ is a 1-form with
values in $\gg$, its field strength or curvature is
the 2-form $F:=\de A+{\textstyle\frac{1}{2}}[A,A]$
again with values in $\gg$. To construct the action
which is a real number, we take the scalar products of
the field strength with itself. The first scalar product
involves the space-time metric $g$ hidden in the Hodge
star $*$, $(\omega,\kappa):=\int_M\omega^**\kappa$,
$\omega$ and $\kappa$ differential forms of same
degree. In components, e.g. for 2-forms
$\omega={\textstyle\frac{1}{2}}\omega_{\mu\nu}\de
x_\mu\de x^\nu$, we have
$(\omega,\kappa)={\textstyle\frac{1}{2}}\int_M
\omega^*_{\mu\nu}\kappa_{\mu'\nu'}g^{\mu\mu'}
g^{\nu\nu'} (\det g_{\cdot\cdot})^{1/2}\,\de^4x$. We
suppose
$M$ Euclidean, otherwise this scalar product would
only be a pseudo scalar product. The second scalar
product is  on the Lie algebra. It only exists if the Lie
group is compact. E.g. for $G=SU(n)$, the general
invariant scalar product is
$(a,b)={\textstyle\frac{2}{g_n^2}}\t (a^*b)$, $a,b\in
su(n)$ and the coupling
constant $g_n$ is a positive number. In general, the
space of all scalar products is a cone whose
coordinates are the coupling constants.

\subsection{The axioms}

Noncommutative geometry does to space-time $M$, a
Riemannian manifold, what quantum mechanics did
to phase space. An uncertainty relation is
introduced by allowing the commutative algebra of
functions
$\ccc^\infty(M)$ to become noncommutative. Let us
call $\aa$ this new algebra that we still suppose
real, associative and equipped with a unit and an
involution.  On phase space, $\aa$ was just the algebra
of observables. Now we want to define a distance on
this new space that has lost its points. Following
Connes, we need a faithful representation $\rho$ of
$\aa$ via bounded operators on a Hilbert space $\hh$,
the space of fermions, and a selfadjoint `Dirac'
operator
$\dd$ on
$\hh$. Connes calls these three ingredients a spectral
triple,
$(\aa,\hh,\dd)$. They satisfy axioms that are
simply taken from the properties of the commutative
case,
$\aa=\ccc^\infty(M)$, the Hilbert space $\hh$ is the
space of ordinary, square integrable Dirac spinors. An
element $f$ of $\aa$ is a differentiable function on
space-time, $f(x)$, and it acts on a spinor $\psi(x)$ by
multiplication $(\rho(f)\psi)(x):=f(x)\psi(x)$.
$\dd=\ddd$ is the ordinary Dirac operator.
Only recently Connes has completed the
list of axioms
\cite {grav} as to have a one-to-one correspondence
between commutative spectral triples and Riemannian
spin manifolds. To this end, he needed two other old
friends from particle physics, a chirality operator
$\chi$ and a real structure $J$. The chirality is a
unitary operator of square one that commutes with the
representation. Therefore $\chi$ decomposes the
representation space
 into a left-handed piece $\frac{1-\chi}{2}\,\hh$ and a
right-handed piece $\frac{1+\chi}{2}\,\hh$. In the
commutative case, of course $\chi=\gamma_5.$ The
real structure is an anti-unitary operator that in the
commutative case reduces to the charge conjugation
operator $C$. $J$ is of square plus or minus one,
depending on space-time
 dimension and signature. Also depending on
space-time dimension and signature, $J$ commutes or
anticommutes with $\chi$. 
The charge conjugation as well decomposes the
representation space into two pieces, particles and
anti-particles, all together
\bb \hh=\hh_L\op\hh_R\op\hh_L^c\op\hh_R^c.
\label{decom}\ee
Here are a few more properties from the commutative
case that become axioms
\begin{itemize}
\item
$\rho(a)$ commutes  with $J\rho(\tilde a)J^{-1}$, 
for all $a,\tilde a \ \in\ \aa,$
\item
$ \dd\chi=-\chi\dd,$ 
\item
$\dd J=+J\dd,$
\item
$[\dd,\rho(a)]$ is bounded for all $a$ in $\aa$,
\item
$[\dd,\rho(a)]$ commutes with  $J\rho(\tilde a)J^{-1}$,
for all
$a,\tilde a$ in $\aa$. 
\end{itemize}
The last axiom is called first order, because in the
commutative case, it just says that the Dirac operator is
a first order differential operator. The dimensionality
of $M$ can be recovered from the spectrum of the
Dirac operator. Indeed for compact manifolds, the
spectrum is discrete and the ordered eigenvalues $\lambda_n$
grow like $n^{1/\dim M}$. This motivates the name
spectral triple. Let us mention two more axioms. The
orientability axiom relates the chirality to the volume
form, a differential form of maximal degree. The
Poincar\'e duality on manifolds is promoted to an
axiom in quite an abstract form. We anticipate that, in
the case of the standard model, this Poincar\'e duality
will prohibit right-handed neutrinos \cite{tresch}.

\subsection{Differential forms}

Our next aim is to construct differential forms starting
from a spectral triple. In the commutative case, we
want this construction to reproduce de Rham's
differential forms, $\Omega M$. 

 We start with an auxiliary differential algebra 
$\Omega \aa,$
the universal differential envelope of $\aa$:
$ \Omega^0\aa := \aa$.   
$\Omega^1\aa$ is
generated by symbols $\delta a$, $ a \in \aa$ with
relations 
$\delta 1 = 0,$   
$\delta(aa') = (\delta a)a'+a\delta a'.$
 $\Omega^1\aa$ consists
of finite sums of terms of the form $a_0\delta a_1,$ and
likewise for higher degree $p$,
\bb   \Omega^p\aa = \left\{ \sum_j
a^j_0\delta a^j_1...\delta a^j_p ,\quad a^j_q\in
\aa\right\}.\eee   
 The differential $\delta$ is defined by
\bb   \delta(a_0\delta a_1...\delta a_p) := 
   \delta a_0\delta a_1...\delta a_p.\eee   
The involution $^*$ is
extended from the algebra $\aa$ to 
 $\Omega^1\aa$ by putting
$(\delta a)^* := \delta(a^*) =:\delta a^*$ and to the
entire differential envelope by
$(\varphi\psi)^*=\psi^*\varphi^*$. 
The next step is to extend the
representation $\rho$  from the  algebra $\aa$   to
its envelope $\Omega\aa$. This
extension deserves a new name: 
\bb  \pi :
\Omega\aa  \longrightarrow \bigoplus_p {\rm
End}(\hh),
   \eee 
 \bb\pi(a_0\delta a_1...\delta a_p) :=
(-i)^p\rho(a_0)[\dd,\rho(a_1)] ...[\dd,\rho(a_p)]
\label{pi}.\eee
$\pi$ is a representation of $\Omega\aa$ as graded
involution algebra, and we are tempted to define also a
differential, again denoted by $\delta$,
 on $\pi(\Omega\aa)$ by
$\delta\pi(\hat\varphi):=\pi(\delta\hat\varphi).$
However, this definition does not make sense because
there are forms
$\hat\varphi\in\Omega\aa$ with
$\pi(\hat\varphi)=0$ and  $\pi(\delta\hat\varphi)
\not= 0$. By dividing out these unpleasant forms,
we arrive at the desired differential algebra
$\Omega_\dd\aa$,
\bb   \Omega_\dd\aa :=
{{\pi\left(\Omega\aa\right)}\over {\cal J}},\qq {\rm
with}
\qq {\cal J} := \pi\left(\delta\ker\pi\right) =:
\bigoplus_p {\cal J}^p,\eee
(${\cal J}$ for junk).  
On the quotient,
the differential is now well defined. Degree by
degree we have:  \bb   \Omega_\dd^0\aa =\rho(\aa)\eee   
because ${\cal J}^0=0$ , 
\bb   \Omega_\dd^1\aa = \pi(\Omega^1\aa)\eee   
because $\rho$ is faithful,
and in degree $p\geq2$ 
\bb   \Omega_\dd^p\aa = 
{{\pi(\Omega^p\aa)}\over
{\pi(\delta(\ker\pi)^{p-1})}}.\eee  
In the commutative case, $\delta=\de$,
$\Omega_\ddd\ccc^\infty(M)
$ is isomorphic to de Rham's differential algebra
$\Omega M$ with
\bb\pi(f_0\de f_1\de f_2...\de f_p)\cong(-i)^p
f_0\gamma^
{\mu_1}\left(\frac{\partial f_1}{\partial
x^{\mu_1}}\right)\gamma^
{\mu_2}\left(\frac{\partial f_2}{\partial
x^{\mu_2}}\right)...\gamma^
{\mu_p}
\left(\frac{\partial f_p}{\partial
x^{\mu_p}}\right).
\label{iso}\ee
Dividing out the junk renders the lhs graded
commutative. The orientability axiom alluded to above
is motivated from this isomorphism, $\de x^1
\de x^2\de
x^3\de x^4\cong(\det
g_{\cdot\cdot})^{1/2}
\gamma^1\gamma^2\gamma^3\gamma^4=
(\det g_{\cdot\cdot})^{1/2}\gamma_5.$

\subsection{The scalar products in noncommutative
geometry}

To play the Yang-Mills game, we need a scalar product
for differential forms. In the noncommutative
context, the scalar product has another utility. It
allows us to interpret the differential forms in
$\Omega_\dd\aa$  not as classes but as concrete
operators on the Hilbert space $\hh$: degree by
degree, we embed
$\Omega^p_\dd$ in $\pi(\Omega^p\aa)$ as orthogonal
complement of $\jj^p$. If $\hh$ was
finite dimensional, we would naturally take as scalar
product of two operators $\omega$ and $\kappa$, 
$<\omega,\kappa>=\t(\omega^*\kappa)$. For infinite
dimensional Hilbert spaces, we have to regularize and
 we use the Dirac operator to do so. Thanks to the
asymptotic behavior of its spectrum, $\t[\omega^*
\kappa\,|\dd|^{-\dim}]$ only diverges logarithmically.
The Dixmier trace ${\rm tr_{Dix}}$ gets rid of this divergence
\cite{dix} and we have a natural scalar product:
\bb <\omega,\kappa>={\rm Re\  tr_{Dix}}[\omega^*
\kappa\,|\dd|^{-\dim}],\qq\omega,\kappa\in
\pi(\Omega^p\aa).\eee
We denote by $(\cdot,\cdot)$ its restriction to
$\Omega_\dd\aa$. In the commutative case of  a four
dimensional manifold $M$, these scalar products are
independent of $M$:
\bb <\omega,\kappa>&=&{\textstyle\frac{1}{32\pi^2}}{\rm Re}
\int_M\t_4[\omega^*\kappa]\de^4x,\qq\omega,\kappa\in
\pi(\Omega^p\aa),\nonumber \\
(\omega,\kappa)&=&{\textstyle\frac{1}{8\pi^2}}{\rm Re}
\int_M\omega^**\kappa,\qq\omega,\kappa\in
\Omega^p_\dd\aa,\eee
where we have used the isomorphism (\ref{iso}) and
view the quotient by the junk as subspace orthogonal
to the junk. We anticipate that this scalar product will
also induce the one we need on the Lie algebra. In
order to get the coupling constants, we soften the
scalar products to:
\bb <\omega,\kappa>_z&=&{\rm Re\ }{\rm tr_{Dix}}[z\omega^*
\kappa\,|\dd|^{-\dim}],\qq\omega,\kappa\in
\pi(\Omega^p\aa)\\
(\omega,\kappa)_z&=&{\rm Re\ }{\rm tr_{Dix}}[z\omega^*
\kappa\,|\dd|^{-\dim}],\qq\omega,\kappa\in
\Omega^p_\dd\aa.\ee
$z$ is a positive operator on Hilbert space that
commutes with $\rho$, $J\rho J^{-1}$, $\dd$ and
$\chi$. Whether or not $z$ commutes with $J$ will be a
difficult choice. In the commutative case, we have
anyhow that $z$ is proportional to the identity.

\subsection{The commutative Yang-Mills action}

The message of this subsection is that the commutative
spectral triple of space-time $M$ is a natural tool to
reconstruct Maxwell's theory: this reconstruction
unifies space-time with internal space, $G=U(1)$. The
first sign for this unification comes from the group of
unitaries of $\aa$. Remember that $\aa$ is the algebra
of complex valued function on $M$ with involution
just complex conjugation. The group of unitaries
$U(\aa):=\{u\in\aa,\ uu^*=u^*u=1\}$ for this algebra
is the group of functions from space-time into $U(1)$
and this is Maxwell's gauge group. Maxwell's four
potential $A\in\Omega_\ddd^1\aa$ is an
anti-Hermitean 1-form on which a gauge
transformation or unitary
$u=\exp i\Lambda$ acts affinely by
\bb
A^u:=\rho(u)A\rho(u^{-1})+\rho(u)\de\rho(u^{-1})
=A-i\de\Lambda.\eee
The field strength
\bb F:=\de A+A^2=\de A\qq\in\Omega_\ddd^2\aa\eee
transforms homogeneouly under unitaries
and is even gauge invariant in the commutative case, 
\bb F^u=\rho(u)F\rho(u^{-1})=F.\eee
The positive operator $z$ from the commutant, that
defines the scalar product can only be a multiple of
the identity. Finally the obviously gauge invariant
Maxwell's action can be written,
\bb S_{\rm Maxwell}[A]&=&(F,F)={\rm Re\ }{\rm tr_{Dix}}( z F^*F|\ddd|^{-4})=
{\textstyle\frac{1}{8\pi^2}}\int_M zF^**F\cr &=&
{\textstyle\frac{z}{16\pi^2}}\int_M F_{\mu\nu}^*
F^{\mu\nu}(\det g_{\cdot\cdot})^{1/2}\ \de^4x=:
{\textstyle\frac{\epsilon_0}{4e^2}}\int_M F_{\mu\nu}^*
F^{\mu\nu}(\det g_{\cdot\cdot})^{1/2}\ \de^4x.\eee
Therefore $z=\pi\hbar c/\alpha_{\rm em}$ with the
fine-structure constant $\alpha_{\rm em}:=e^2/(4\pi
\epsilon_0\hbar c)$. Later, we will call $z$
noncommutative coupling constant. Had we dropped
the condition that
$z$ commute with the Dirac operator, we would have
inherited an $x$ dependent coupling `constant'. 

The commutative pure Yang-Mills theory is linear and
to justify the word coupling constant, we have to add
matter, say an electron $\psi$. The Dirac operator acts
on it defining its kinetic energy, unitaries act on it by
\bb \psi^u=\rho(u)\psi,\qq u\in U(\aa),\qq
\psi\in\hh, \eee
and we define the minimal coupling by the covariant
Dirac operator
$\ddee:=\ddd+\pi(A)$. The Dirac action then reads
\bb S_{\rm Dirac}[\psi,A]=\int_M\psi^*\ddee\psi\,
(\det g_{\cdot\cdot})^{1/2}\ \de^4x,\eee
where here the star denotes the dual with respect to
the scalar product of the Hilbert space $\hh$. A mass
term $m_\psi\psi^*\psi$ may be added.

Let us stress again that in Connes' formulation, the
gauge coupling, that is the invariant scalar
product in internal space, is induced from the scalar
product of differential forms over space-time.

\subsection{The tensor product}

One way to see the above commutative example is to say
that the associative algebra of the spectral triple is
$\aa_t=\ff\ot\aa_f$, a tensor product of the
 commutative, infinite dimensional algebra of
 {\it real} valued functions $\ccc^\infty(M)$ on
space-time and the commutative, finite dimensional,
{\it real} algebra
$\aa_f=\cc$. The gauge group then is Abelian,
$G=U(1)\subset\aa_f$. It is natural to try
noncommutative algebras for $\aa_f$ to get
non-Abelian gauge groups. In this spirit, we have to
consider tensor products of entire spectral triples, and
the message of this subsection is that if the
fermionic representation breaks parity, the Higgs
scalar and the symmetry breaking potential come free
of charge.

Let us denote by $(\ff,\sss,\ddd,\gamma_5,C)$
the commutative spectral triple of a four dimensional
space-time and by
$(\aa_f,\hh_f,\dd_f,\chi_f,J_f)$, $\cdot_f$ for finite,
the one of a (zero dimensional) internal space. Note
that our  $C$ is anti-unitary.
According to the rules of noncommutative geometry
the tensor product of these two spectral triples
$(\aa_t,\hh_t,\dd_t,\chi_t,J_t)$, $\cdot_t$ for tensor,
is:
\bb
\aa_t\,=\,\ff\ot\aa_f,\qq
&\hh_t\,=\,\sss\ot\hh_f,\qq&
\dd_t\,=\,\ddd\ot 1\,+\,\gamma_5\ot\dd_f,\cr 
&\chi_t\,=\,\gamma_5\ot\chi_f,\qq &J_t\,=\,C\ot
J_f.\eee 
Before turning the crank, we must talk about
the internal Dirac operator $\dd_f$. From the axioms,
we infer that with respect to the decomposition
(\ref{decom}) of the fermionic Hilbert space $\hh_f$
the internal Dirac operator has the form:
\bb \dd_f=\pp{0&\mm&0&0\cr 
\mm^*&0&0&0\cr 
0&0&0&\overline{\mm}\cr 
0&0&\overline{\mm^*}&0},\eee
where $\mm$ is the fermionic mass matrix. This is
another manifestation of the unification of space-time
and internal space, the naked Dirac operator $\ddd$
and its mass matrix satisfy the same list of axioms above. 

As in the commutative case, we start by identifying
the gauge group, the functions from space-time into
the finite dimensional Lie group $G=U(\aa_f)$.  It is
represented affinely on the bosonic fields. They are
anti-Hermitean 1-forms.  But now,
\bb \Omega^1_{\dd_t}\aa_t=
\Omega^1_\ddd\ff\ot\Omega^0_{\dd_f}\aa_f\,\op\,
\Omega^0_\ddd\ff\ot\Omega^1_{\dd_f}\aa_f\,\cong\,
\Omega^1(M,\aa_f)\,\op\,
\ff\ot\Omega^1_{\dd_f}\aa_f\ \owns A_t=:A \oplus H.\eee
From the anti-Hermiticity of $A_t$, it
follows that
$A$ is in fact a Lie algebra valued 1-form on space-time,
$A\in\Omega^1(M,\gg)$, i.e. a Yang-Mills potential.
$\gg:=u(\aa_f):=\{X\in\aa_f,\ X+X^*=0\}$ is the Lie
algebra of the group of unitaries
$G=U(\aa_f)$.  On the other hand,
 the Higgs scalar $H$ is a 0-form on space-time, valued
in a representation of the Lie group
$G$. The inhomogeneous transformation law,
\bb A_t^u&=&
\rho_t(u)A_t\rho_t(u^{-1})+\rho_t(u)\de_t\rho_t(
u^{-1}) =A^u \oplus H^u,\cr &&
A^u=uAu^{-1}+u\de u^{-1},\qq
H^u=\rho_f(u)H\rho_f(u^{-1})+
\rho_f(u)\delta\rho_f(u^{-1}),
\eee
determines according
to which {\it group} representation the Higgs scalar
transforms and this depends on the details of the
internal spectral triple. We denote by $\rho_t$ the
representation of
$\aa_t$ on $\hh_t$, by $\rho_f$ the representation of
$\aa_f$ on $\hh_f$, by $\delta_t$ the differential of
$\Omega_{\dd_t}\aa_t$ and so forth.
Next we define the field strength,
\bb F_t =
\delta_tA_t+A_t^2\qq\in\Omega^2_{\dd_t}\aa.\eee
To decompose the field strength, it is comfortable to
change scalar variables,
\bb \Phi(x)\,:=\,H(x)-i\dd_f\,=\,-\Phi^*(x)\ \in
\Omega^0(M,\Omega^1_{\dd_f}\aa_f).\eee 
This change of variables is well defined within
$\Omega^0(M,\Omega^1_{\dd_f}\aa_f)$ thanks to the
orientability axiom \cite{tk}. $\Phi$ has the good
taste to transform homogeneously under a gauge
transformation $u$ and we can define its covariant
exterior derivative, 
\bb\dee\Phi:= \de\Phi+\left[\rho_f(A),\Phi\right]
\ \in\Omega^0(M,\Omega^1_{\dd_f}\aa_f).\eee
The field strength decomposes as
\bb F_t=F+(C-\alpha C)-\dee\Phi\gamma_5),\eee
with
\bb F&=&\de A+A^2\ \in\Omega^2(M,\gg), \nonumber \\
C&=&\delta H+H^2
\in\Omega^0(M,\Omega^2_{\dd_f}\aa_f).\eee
The
internal field strength $C$, for curvature, should not
be confused with the $C$ of charge conjugation. 
$\alpha C\in\Omega^0(M,\Omega^2_{\dd_f}\aa_f+
{\cal J}^2_f)$ is the tricky piece of the computation, it
comes from the interference in degree two of
space-time junk and internal junk. The former is
isomorphic to $\Omega^0 M$, a first happy
circumstance. A second is that the positive operator
$z_t$ in the scalar product is necessarily of the form
$z_t=1\ot z_f$. Both circumstances together allow to compute 
$\alpha C$ pointwise \cite{sz}. For fixed $x$,
$C\in\Omega_{\dd_f}\aa_f\subset {\rm End}\hh_f$
and
$\alpha C\in
\pi(\Omega\aa_f)\subset {\rm End}\hh_f$ are finite
dimensional operators, i.e. matrices. 

Let us denote by 
$<\omega,\kappa>_{z_f}\ =\re\t [z_f\omega^*\kappa]$,
the finite dimensional scalar product. Then $\alpha C$
is uniquely determined by the linear equations
\bb <r,C-\alpha C>_{z_f}\ =0&\qq {\rm for\ all}\ &
r\in\rho_f(\aa_f),\label{r}\\ 
<j,C-\alpha C>_{z_f}\ =0&\qq {\rm for\ all}\ &
j\in\jj^2_f,\label{j}\ee
where the trace is over the finite dimensional Hilbert
space $\hh_f$. 
Under a gauge transformation $u(x)$, the field
strength transforms homogeneously and we can
define, as before, the Yang-Mills action,
\bb S_{\rm YM}[F_t]=(F_t,F_t)_{z_t} = {\rm Re\ }{\rm tr_{Dix}}( z_t
F_t^*F_t|\dd_t|^{-4}).\eee
The differential algebra
contains the Lie algebra as 0-forms and the scalar
product $(\cdot,\cdot)_{z_t}$ with $z_t=1\ot z_f$
 restricted to the Lie algebra is an
invariant scalar product. Therefore this action is
gauge invariant. Let us decompose it,
$S_{\rm YM}[F_t]=S_{\rm YM}[F,H]$: 
\bb S_{\rm YM}[F,H]=
{\textstyle\frac{1}{8\pi^2}}\int_M< F,*F>_{z_f}+\,
{\textstyle\frac{1}{8\pi^2}}\int_M< \dee\Phi,*
\dee\Phi>_{z_f}+\,
{\textstyle\frac{1}{8\pi^2}}\int_M*V(H),\eee
with
\bb V(H)=\ <C-\alpha C,C-\alpha C>_{z_f}\ 
=(C,C)_{z_f}\,-<\alpha C,\alpha C>_{z_f}.\eee
The first term, a non-Abelian Yang-Mills action, is no
surprise. The second, a Klein-Gordon action,
propagates the Higgs scalar. The Higgs potential
$V(H)$ breaks the gauge group spontaneously, if the
fermions break parity. As we shall see, the
computation of the Higgs sector, representation and
potential, will be intricate even though it follows from
a simple geometric definition,
$S_{\rm YM}[F_t]=(F_t,F_t)_{z_t}$. This simple
geometric definition constrains the ensuing
Yang-Mills theory. The Lie group $G$ is not arbitrary,
it must be a group of unitaries of an associative
algebra, which is not the case for the exceptional
groups. Furthermore, the fermionic representation is
not only a representation of the group but must also
be a representation of the algebra which is not the
case for representations other than the fundamental
ones. Finally, the Higgs representation is computed,
not chosen. In any case, no left-right symmetric and
no grand unified theory admits a formulation within
noncommutative geometry.

To end this subsection, we mention the Dirac
Lagrangian, $\lll_{\rm Dirac}=\psi^*\dd_{t,{\rm
cov}}\psi$. The total, covariant Dirac operator is
\bb \dd_{t,{\rm cov}}=\dd_t+\pi_t(A_t)+
J_t(\dd_t+\pi_t(A_t))J_t^{-1}.\label{covdir}\ee
Note the appearance of charge conjugation that
will be crucial. The decomposition of this Lagrangian
is:
\bb\lll_{\rm
Dirac}=\psi^*(\ddd-i\rho_f(\aaa)-J_ti\rho_f(\aaa)
J_t^{-1})
\psi-
\psi^*(\Phi\gamma_5+J_t\Phi\gamma_5J_t^{-1})\psi.\eee 
In words: noncommutative geometry promotes the
Higgs scalar to a connection and thereby unifies the
gauge couplings hidden in $\rho_f(\aaa)$ with
the Yukawa couplings hidden in $\Phi$.

\subsection{The standard model}
\subsubsection{The algebraic setting}
It is time for an example. Concerning its choice, we
emphasize two points. The standard model has an
internal space that does fit the elaborate axioms of a
spectral triple. The internal spectral triple of the
standard model is not far from being the simplest,
non-degenerate example. To make this more precise,
we note that the standard model viewed as an ordinary
Yang-Mills-Higgs theory has the following four {\it
unrelated} features: 

(i) weak interactions break parity
maximally, 

(ii) weak interactions suffer spontaneous
break down, 

(iii) strong interactions do not break
parity, 

(iv) strong interactions do not suffer
spontaneous break down. \\
Flip just one of these 
features and the standard model is outside the noncommutative axioms
\cite{becca}\cite{tk}.
A more quantitative constraint concerns the Higgs
representation, that in Connes' formulation is not
chosen but computed. The spectral triple of the
standard model implies that the Higgs scalar
transforms as one doublet under weak isospin
entailing a unit $\rho$-factor,
\bb
\rho:=\frac{m_W^2}{\cos^2(\theta_w) \ m_Z^2}=1.
\eee
Experimentally we have today $\rho=1.0012\pm
0.0031$. 

The geometric version of the standard model is well
documented in the literature
\cite{book}\cite{cl}\cite{tresch}\cite{reviews} and we
just have to fix our notations. We denote by $\hhh$ the
algebra of quaternions, viewed as $2\times 2$
matrices,
\bb \pp{x&-\bar y\cr y&\bar x}\,\in\hhh,\qq
x,y\in\cc.\eee
From now on, everything concerns
the internal spectral triple and we drop the subscript
$f$ for finite,
\bb \aa&=&\hhh\op\cc\op M_3(\cc)\,\owns\,(a,b,c), \nonumber \\
\hh_L&=&
\left(\cc^2\ot\cc^N\ot\cc^3\right)\ \op\ 
\left(\cc^2\ot\cc^N\ot\cc\right),\nonumber \\
\hh_R&=&\left((\cc\op\cc)\ot\cc^N\ot\cc^3\right)\ 
\op\ \left(\cc\ot\cc^N\ot\cc\right).\eee
 In each summand, the first factor
denotes weak isospin doublets or singlets, the second
$N$ generations, $N=3$, and the third denotes color
triplets or singlets.
Let us choose the following basis
of 
$\hh=\cc^{90}$: 
\bb
& \pp{u\cr d}_L,\ \pp{c\cr s}_L,\ \pp{t\cr b}_L,\ 
\pp{\nu_e\cr e}_L,\ \pp{\nu_\mu\cr\mu}_L,\ 
\pp{\nu_\tau\cr\tau}_L;&\cr \cr 
&\matrix{u_R,\cr d_R,}\qq \matrix{c_R,\cr s_R,}\qq
\matrix{t_R,\cr b_R,}\qq  e_R,\qq \mu_R,\qq 
\tau_R;&\cr  \cr 
& \pp{u\cr d}^c_L,\ \pp{c\cr s}_L^c,\ 
\pp{t\cr b}_L^c,\ 
\pp{\nu_e\cr e}_L^c,\ \pp{\nu_\mu\cr\mu}_L^c,\ 
\pp{\nu_\tau\cr\tau}_L^c;&\cr\cr  
&\matrix{u_R^c,\cr d_R^c,}\qq 
\matrix{c_R^c,\cr s_R^c,}\qq
\matrix{t_R^c,\cr b_R^c,}\qq  e_R^c,\qq \mu_R^c,\qq 
\tau_R^c.&\eee
The representation $\rho$ acts on $\hh$ by
\bb \rho(a,b,c):=\pp{\rho_{w}(a,b)&0\cr 0&\rho_{s}(b,c)} := 
\pp{\rho_{wL}(a)&0&0&0\cr 
0&\rho_{wR}(b)&0&0\cr 
0&0&\overline{\rho_{sL}(b,c)}&0\cr 
0&0&0&\overline{\rho_{sR}(b,c)}}\eee
with
\bb\rho_{wL}(a)&:=&\pp{
a\ot 1_N\ot 1_3&0\cr
0&a\ot 1_N&},\qq
\rho_{wR}(b)\ :=\ \pp{
B\ot 1_N\ot 1_3&0\cr
0&\bar
b1_N},\cr &&
B:=\pp{b&0\cr 0&\bar b},
\nonumber \\   
  \rho_{sL}(b,c)&:=&\pp{
1_2\ot 1_N\ot c&0\cr
0&\bar b1_2\ot 1_N},\qq
\rho_{sR}(b,c)\ :=\ \pp{
1_2\ot 1_N\ot c&0\cr
0&\bar b1_N}.   
\eee
The chosen representation $\rho$ will take into
account weak interactions $\rho_w(a,b),\ a\in\hhh,\ 
b\in\cc$, and strong interactions $\rho_s(b,c),\ c\in
M_3(\cc)$, $c$ for color. This choice discriminates
between leptons (color singlets) and quarks (color
triplets). The role of $b\in\cc$ appearing in both
weak interactions $\rho_w(a,b)$ and strong
interactions $\rho_s(b,c)$ is crucial to make
$\rho(a,b,c)$ a representation of $\aa$ and is crucial
for weak hypercharge computations. There is an
apparent asymmetry between particles and
anti-particles, the former are subject to weak, the
latter to strong interactions. However, since particles
and anti-particles are permuted in the covariant Dirac
operator (\ref{covdir}) by
\bb J=\pp{0&1_{15N}\cr 1_{15N}&0}\circ c.c.,\eee
the theory is invariant under charge conjugation.
We denote the complex conjugation by $c.c.$. For
completeness, we record the chirality as matrix
\bb \chi=\pp{-1_{8N}&0&0&0\cr 0&1_{7N}&0&0\cr 0&0&-1_{8N}&0\cr 0&0&0&1_{7N} }.\eee
The
third item in the spectral triple is the Dirac operator
\bb \dd=\pp{0&\mm&0&0\cr 
\mm^*&0&0&0\cr 
0&0&0&0\cr 
0&0&0&0}.\eee
The fermionic mass matrix of the standard model is
\bb\mm=\pp{
\pp{M_u&0\cr 0&M_d}\ot 1_3&0\cr
0&\pp{0\cr M_e}},\eee
with
\bb M_u:=\pp{
m_u&0&0\cr
0&m_c&0\cr
0&0&m_t},\qq M_d:= C_{KM}\pp{
m_d&0&0\cr
0&m_s&0\cr
0&0&m_b},\qq M_e:=\pp{
m_e&0&0\cr
0&m_\mu&0\cr
0&0&m_\tau}.\eee
 All indicated fermion masses are supposed positive and
different. The
Cabibbo-Kobayashi-Maskawa matrix  $C_{KM}$
is supposed non-degenerate in the sense that there is
no simultaneous mass and weak interaction
eigenstate. Note that the
 strong interactions are vector-like and $\rho_s$
commutes with $\dd$.

Let us compute the noncommutative coupling constant
$z$. We recall that $z$ is a positive operator on $\hh$ that 
commutes with the representation $\rho$, with its
opposite $J\rho J^{-1}$, with the chirality $\chi$, and
with the Dirac operator $\dd$. It follows that $z$
involves 
$2(1+N)=8$
strictly positive numbers $x,\ y_1,\ y_2,\ y_N,$ $ \tilde
x,\ \tilde y_1,\  \tilde y_2,\ \tilde y_N$,
\bb 
z&:=& \pp{z_w&0\cr 0& z_s},\nonumber \\
z_w&:=&\pp{
x/3\,1_2\ot 1_N\ot 1_3&0&0&0\cr  0&1_2\ot y&0&0\cr 
0&0&x/3\,1_2\ot 1_N\ot 1_3&0\cr 
0&0&0&y},\cr \cr \cr 
z_s&:=&\pp{
\tilde x/3\,1_2\ot 1_N\ot 1_3&0&0&0\cr  0&1_2\ot 
\tilde y&0&0\cr 
0&0&\tilde x/3\,1_2\ot 1_N\ot 1_3&0\cr 
0&0&0&\tilde y},\cr\cr  \cr 
 y&:=&\pp{ 
y_1&0&0\cr 
0&y_2&0\cr 
0&0&y_N},\qq 
\tilde y:=\pp{ 
\tilde y_1&0&0\cr 
0&\tilde y_2&0\cr 
0&0&\tilde y_N}.\eee
The interpretation of these numbers is
straightforward. The three $y_j$ poise the weak
interactions with the three lepton generations. The
$y_j$ enter independently because the Higgs scalar
couples differently to the three leptons and in
noncommutative geometry the Higgs is part of
the gauge
interactions.  The three $\tilde y_j$ poise the `strong'
interactions with the three lepton generations. They
do not drop out because of the $b$ in $\rho_s$.
However, they will only enter as sum: strong
interactions are unbroken and do not generate a
Higgs. We will denote $\tilde y:=\tilde y_1+\tilde
y_2+\tilde y_N$ and there should be no risk of
confusion.
$x$ and $\tilde x$ poise weak and strong interactions
with quarks. There is only one number per interaction
because of the Cabibbo-Kobayashi-Maskawa mixing
that we suppose non-degenerate.

\subsubsection{The choice of a scalar product}
We recall the internal scalar product
$<\omega,\kappa>_z\ =\re\t [z\omega^*\kappa]$,
$\omega,\
\kappa\in\pi(\Omega\aa)$. At this point comes the
new lesson from noncommutative relativity. 
It tells us that we have forgotten an entire
cone of other scalar products,
\bb 
<\omega,\kappa>_{z'}\  :=\ \re\t [z'(\omega+J\omega J^{-1})^*
(\kappa+J\kappa J^{-1})]
\eee
 with additional $1+N$
strictly positive constants $x',\ y_1',\ y_2',\ y_N'$,
\bb 
z'&:=& \pp{z'_w&0\cr 0&z'_s},\nonumber \\
z'_w&=&z'_s:={\textstyle\frac{1}{2}}\pp{
x'/3\,1_2\ot 1_N\ot 1_3&0&0&0\cr  0&1_2\ot y'&0&0\cr 
0&0&x'/3\,1_2\ot 1_N\ot 1_3&0\cr 
0&0&0&y'}.\eee
Indeed, in noncommutative relativity, the scalar product is not chosen, it is induced
from the heat kernel calculation. The Dirac operator 
$\dd_{t,{\rm cov}}=\dd_t+\pi_t(A_t)+
J_t\pi_t(A_t)J_t^{-1}$ leads to the scalar product
with $z'$. We could obtain the one with $z$ from
another Dirac operator, $\dd_{t,{\rm cov}}=
\dd_t+\pi_t(A_t)$, but this latter is forbidden by the
spectral principle: for a unitary $u \in \aa_t$, the inner automorphim 
\bb
\alpha_u :\rho_t(a) \in \rho_t(\aa_t) \longrightarrow \rho_t(uau^*) \in \rho_t(\aa_t)
\eee
induces a unitary operator $U=uJuJ^{-1}$ on $\hh_t$ satisfying
\bb
U\rho_t(a)U^* = \alpha_u(\rho_t(a)), \ {\rm and} 
\ U \dd_t U^*= \dd_t + A + JAJ^{-1} \ {\rm with} \ 
A=u[\dd_t,u^*],
\eee
so $\dd_t$ and $\dd_{t,cov}$ have the same spectrum and 
the fluctuations of the metric are of the form $A + JAJ^{-1}$.

\begin{figure}[hbt]
\hspace{2cm}
\def\epsfsize#1#2{0.6#1}
\epsfbox{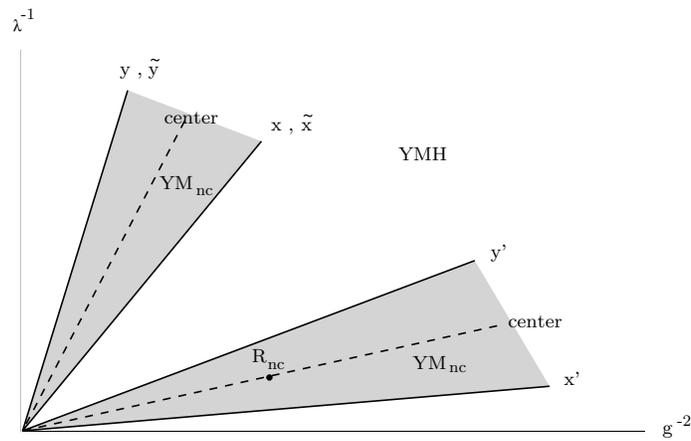}
\caption{The allowed cones of possible scalar products}
\label{fig:The allowed cones}
\end{figure}

\noindent Restricted to the Lie algebra $\gg$, we have a subtle
nuance between the two invariant scalar products
concerning the two $u(1)$ factors. For $z'$ in the
center $\rr^+1$, we have
$\sin^2\theta_w=3/8$, while for $z$ in the center we
will get 
$\sin^2\theta_w=12/29$. Note also that $z'$ commutes
with the real structure $J$, while $z$ does not. If in
doubt, stay out: we will use both cones simultaneously.
Figure \ref{fig:The allowed cones} is an artist's view on 
the role of the possible scalar products in 
noncommutative Yang-Mills theory:
\bb<\omega,\kappa>_{z,z'}\ :=\re\t
[z\omega^*\kappa]+\re\t
[z'(\omega+J\omega J^{-1})^*
(\kappa+J\kappa J^{-1})],\qq\omega,\
\kappa\in\pi(\Omega\aa). \label{produitscaire}
\ee

\subsubsection{The gauge couplings computation}
We are ready to turn the crank. A long, but straight
down the line computation leads to the physical
couplings in terms of the fermionic mass matrix
$\mm$ and the noncommutative couplings $z,\ z'$:

Here are a few purely algebraic intermediate steps:
\bb   \Omega_\dd^1\aa = \left\{i\pmatrix
{0&\rho_{wL}(h)\mm&0&0\cr \mm^*\rho_{wL}(\tilde
h^*)&0&0&0\cr 
0&0&0&0\cr 0&0&0&0}
,\ h,\tilde h\in \hhh\right\}.\eee
The Higgs being an anti-Hermitian 1-form
\bb H=  i\pmatrix
{0&\rho_{wL}(h)\mm&0&0\cr 
\mm^*\rho_L(h^*)&0&0&0\cr 
0&0&0&0\cr 0&0&0&0},\qq
h=\pp{h_1&-\bar h_2\cr h_2&\bar h_1}\in \hhh\eee
is parameterized by one complex doublet
\bb \pp{h_1\cr h_2},\qq h_1,h_2\in\cc.\eee
The internal junk in degree two turns out to be
\bb \jj^2=\left\{i\pp{j\ot\Delta&0&0&0\cr 
 0&0&0&0\cr 
 0&0&0&0\cr 0&0&0&0},\qq
j\in\hhh\right\}\eee
with
\bb\Delta:={\textstyle\frac{1}{2}}\pp{
\left(M_uM_u^*-M_dM_d^*
\right)\ot 1_3&0\cr 
0&-M_eM_e^*}.\eee
The homogeneous scalar variable is:
\bb \Phi:= H-i\dd=:i\pmatrix
{0&\rho_{wL}(\phi)\mm&0&0\cr
\mm^*\rho_{wL}(\phi^*)&0&0&0
\cr 0&0&0&0\cr 0&0&0&0},\qq
\phi=\pp{\varphi_1&-\bar \varphi_2\cr
\varphi_2&\bar \varphi_1}\in \hhh,\eee
and with
$\varphi:=(\varphi_1,\varphi_2)^T$,
the internal field strength is:
\bb C&:=&\delta H+H^2=
\left(1-|\varphi|^2\right)\pp{1_2\ot\Sigma&0&0&0
\cr 0&\mm^*\mm&0&0
\cr 0&0&0&0\cr 0&0&0&0}, \nonumber \\
&&\Sigma:={\textstyle\frac{1}{2}}\pp{
\left(M_uM_u^*+M_dM_d^*
\right)\ot 1_3&0\cr 
0&M_eM_e^*}.\eee 

Now, the chosen scalar product (\ref{produitscaire}) appears in the long computation of $\alpha C$ from
(\ref{r}) and (\ref{j}). In the standard model,
equation (\ref{j}) implies that
$\alpha C$ has no junk component, and has the form
\bb\alpha C=(1-|\varphi|^2)m_t^2\,\rho(\alpha
1_2,\beta,\gamma 1_3).\eee
To compute the real numbers $\alpha,\ \beta,\
\gamma$, we neglect all fermion masses with respect
to the top mass. This approximation is again
good to $m_b^2/m_t^2= 0.0006$ \cite{cis}. In this approximation,
the number of parameters reduces again, we are left
with six parameters: $x,\ \tilde x,\ y:=y_1+y_2+y_N,\ x',$
and $y':=y'_1+y'_2+y'_N, \tilde y$, and $\alpha,\ \beta,\
\gamma$ are determined by three linear equations:
\bb\begin{array}{rcrcrcr}
[N(x+x')+y+y'] \alpha&+&y'\, \beta&+&Nx'\, \gamma
&=&
{\textstyle\frac{1}{2}}(x+x')\cr 
2y'\, \alpha&+&[2N(x+x')+y+3\tilde y+6y'] \beta&+&
2Nx'\, \gamma&=&\ (x+x')\cr 
Nx'\, \alpha&+&Nx' \,\beta&+&2N(\tilde
x+x') \gamma&=& \qq x'.
\end{array}\label{sys}\ee
The Higgs Lagrangian has the form:
\bb
&&{\textstyle\frac{2}{8\pi^2}}(x+x')m_t^2\,
(\dee_\mu\varphi)^*\dee^\mu\varphi\,+\,
{\textstyle\frac{1}{8\pi^2}}
k\,m_t^4\left(1-|\varphi|^2\right)^2\cr 
&&\qq=:{\textstyle\frac{1}{2}}\,(\dee_\mu\varphi_{\rm
ph})^*
\dee^\mu\varphi_{\rm ph}\,+\,\lambda
|\varphi_{\rm ph}|^4
\,-\,{\textstyle\frac{1}{2}}\,\mu^2|\varphi_{\rm
ph}|^2 + {\rm constant} ,\eee
with
\bb k&:=&{\textstyle\frac{3}{2}}(x+x')-2Nx(\alpha^2
+\beta^2)-y(2\alpha^2+\beta^2)
-4N\tilde x\gamma^2\cr 
&&-3\tilde y\beta^2
-2Nx'((\alpha+\gamma)^2+(\beta+\gamma)^2)
-y'(2(\alpha+\beta)^2+4\beta^2).\eee
Therefore 
\bb\lambda^{-1}&=&{\textstyle\frac{2}{\pi^2}}\frac{(x+x')^2}{k},\label{lam}\\
\mu^2&=&\frac{k}{x+x'}\;m_t^2.\label{mu}\ee
Before
computing the gauge couplings, we have to get rid of
the unwanted
$u(1)$ in $\gg$. This is done by imposing the
unimodularity condition, 
\bb\t \left[P\left(\rho(a,b,c)+J\rho(a,b,c)J^{-1}
\right)\right]=0,\eee
 where $P$ is the projection on $\hh_L\op\hh_R$, the
space of particles. Note
that this condition is equivalent to the condition of
vanishing gauge anomalies \cite{reviews}. Normalizing properly the
gauge fields, we compute their couplings:
\bb g_3^{-2}&=&{\textstyle\frac{1}{6\pi^2}}N(\tilde
x+x'),\label{g3}\\ 
g_2^{-2}&=&{\textstyle\frac{1}{8\pi^2}}[N(x+x')
+y+y'],\label{g2}\\ 
g_1^{-2}&=&{\textstyle\frac{1}{8\pi^2}}
[Nx+{\textstyle\frac{2}{9}}N\tilde
x+{\textstyle\frac{11}{9}}Nx'+
{\textstyle\frac{1}{2}}y+{\textstyle\frac{3}{2}}\tilde y
+3 y'].\label{g1}\ee

\subsection{Results}

As for noncommutative relativity,
we interpret the five constraints
(\ref{lam}-\ref{g1}) in terms
of running quantities at the noncommutative scale
$\Lambda$. Since the flow of $\mu^2$ is
renormalization scheme dependent, we trade the
running top mass for its Yukawa coupling, $m_t=g_tv$,
$m_W={\textstyle\frac{1}{2}}g_2v$, $m_H=2\sqrt {2
\lambda}\; v$,
$v={\textstyle\frac{1}{2}}\frac{\mu}{\sqrt\lambda}$,
\bb
g_3(\Lambda)^{-2}&=&{\textstyle\frac{1}{6\pi^2}}N(\tilde
x+x'),\label{g3L}\\ 
g_2(\Lambda)^{-2}&=&{\textstyle\frac{1}{8\pi^2}}
[N(x+x')+y+y'],\label{g2L}\\ 
g_1(\Lambda)^{-2}&=&{\textstyle\frac{1}{8\pi^2}}
[Nx+{\textstyle\frac{2}{9}}N\tilde
x+{\textstyle\frac{11}{9}}Nx'+
{\textstyle\frac{1}{2}}y+{\textstyle\frac{3}{2}}\tilde y
+3 y'],\label{g1L}\\
\lambda(\Lambda)^{-1}&=&{\textstyle\frac{2}{\pi^2}}\frac{(x+x')^2}{k},\label{lamL}\\
g_t(\Lambda)^{-2}&=&{\textstyle\frac{1}{2\pi^2}}
(x+x').\label{gtL}\ee
These Yang-Mills constraints are to be compared to the
soft Einstein-Hilbert constraints 
\bb
g_3(\Lambda)^{-2}&=&{\textstyle\frac{1}{9\pi^2}}
f_4Nx',
\label{eh3}\\ 
g_2(\Lambda)^{-2}&=&{\textstyle\frac{1}{12\pi^2}}
f_4(Nx'+y'),\\ 
g_1(\Lambda)^{-2}&=&{\textstyle\frac{1}{12\pi^2}}f_4(
{\textstyle\frac{11}{9}}Nx'+3y')
\label{eh1},\\ 
\lambda(\Lambda)^{-1}&=&{\textstyle\frac{1}{\pi^2}}
f_4\,x'.\label{ehlam}\ee
The noncommutative Yang-Mills action has four
additional parameters, $x,\ \tilde x,\ y,\ \tilde y$, but
one additional constraint, on the top mass. 

These results can be detailed at different levels, playing 
with $z$ and $z'$. 

$\bullet$ The original Connes-Lott model
\cite{cl}\cite{book} used $z_w=:z_1$, $z_s=:z_2$, 
$z'=0$ and $\Lambda=m_Z$, i.e. tree level. It worked
with a bimodule and had {\it two} spurious $U(1)$
factors. Consequently its linear system (\ref{sys}) is
slightly different, put $\tilde y=0$, and the Higgs
mass comes out \cite{ks} :
\bb m_H^2&=&
3\,\frac{(m_t/m_W)^4+2(m_t/m_W)^2-1}
{(m_t/m_W)^2+3}\,\ m_W^2,\nonumber \\
m_H&=&278 \ {\rm GeV}, \eee
for $m_t=175$ GeV. 

$\bullet$ With the real
structure \cite{tresch}, we are inflicted with only one
spurious
$U(1)$. If we put $z'=0$, then we can solve the
system (\ref{sys}) even without the approximation of
a dominating top mass:
\bb\alpha={\textstyle\frac{1}{2}}\frac{x}{Nx+y},\qq
\beta={\textstyle\frac{1}{2}}\frac{x}{Nx+
{\textstyle\frac{1}{2}}y+{\textstyle\frac{3}{2}}\tilde
y},\qq\gamma=0,\eee
and at tree level
\cite{cis}:
\bb m_H^2&=&3\, {m_t}^2-\left(1+\frac{{g_2}^
 {-2}}{{g_1}^{-2}-\frac{1}{6}{g_3}^{-2}}
 \right)\,m_W^2  \label{mH},\nonumber \\
m_H&=&289 \ {\rm GeV}\eee
for $m_t=175$ GeV. In this case, we also have precise
results with all fermion masses and mixings. The
$\tau$ mass renders the Higgs mass fuzzy
with a relative uncertainty of the order of
$m_\tau^2/m_t^2$, that is some tens of MeV, as  $y_3$
ranges from 0 to its maximal value.

$\bullet$ Our general analysis including $z$, $z'$ and $\Lambda$
starts with the inequality,
\bb g_2(\Lambda)<{\textstyle\frac{2}{\sqrt N}}
g_t(\Lambda)\ee
coming from equations (\ref{g2L}) and (\ref{gtL}).
Identifying the pole masses of the $W$ and top with
their running masses at $m_Z$, {\it this inequality  sets an
upper bound}  $\Lambda_{\max}$ on the
noncommutative scale shown in Figure 
\ref{fig:The cutoff as function of the top mass}. This bound
is rather sensitive to variations in the gauge
couplings. The noncommutative Einstein-Hilbert
action needs a scale $\Lambda$ of at least $10^{10}$
GeV forcing a high top mass or slightly different
gauge couplings as suggested anyhow by the stiff
action. 

In the presence of $z$, the top mass is a free
parameter and
$z'$ is just a perturbation rendering the Higgs mass
fuzzy. This comes from the fact that $x'$ and $y'$ are
bounded from above,
\bb
x'_{\max}&\! =&\! \min \{ {\textstyle 2\pi^2}
g_3(\Lambda)^{-2},\,{\textstyle 2\pi^2}
g_t(\Lambda)^{-2}   \} \nonumber \\ 
y'_{\max}&\! =&\! \min \{ 
{\textstyle 8\pi^2}g_2(\Lambda)^{-2}-{\textstyle 6\pi^2}
g_t(\Lambda)^{-2},\,
{\textstyle\frac{16\pi^2}{5}}g_1(\Lambda)^{-2}
-{\textstyle\frac{8\pi^2}{5}}g_2(\Lambda)^{-2}
-{\textstyle\frac{8\pi^2}{15}}g_3(\Lambda)^{-2}
-{\textstyle\frac{6\pi^2}{5}}g_t(\Lambda)^{-2} \}
\eee
and that the Higgs mass decreases with $x'$, increases
with  $y'$. With $m_t=175 \pm 6$ GeV, this fuzziness is:
\bb m_H&=&289^{+2}_{-5}\ {\rm GeV \qq for \qq}
\Lambda=m_Z, \cr 
m_H&=&195^{+.0}_{-.5}\ {\rm GeV \qq for \qq}
\Lambda=\Lambda_{\max}=2\cdot 10^5\ {\rm GeV}.\ee
This narrow interval accessible to the Higgs mass
comes of course from a narrow interval accessible to
the scalar selfcoupling, 
\bb\lambda(\Lambda)&=& (0.329\,-\,0.345)\,g_3^2 \qq
{\rm for} \qq\Lambda=m_Z, \cr 
\lambda(\Lambda)&=& (0.313\,-\,0.317)\,g_3^2 \qq
{\rm for} \qq
\Lambda=\Lambda_{\max}=2\cdot 10^5\ {\rm GeV}.\eee
Finally, we have one other constraint in presence of $z$
and $z'$,
\bb 
{\textstyle\frac{1}{15}}g_3(\Lambda)^{-2}+
{\textstyle\frac{1}{5}}g_2(\Lambda)^{-2}+
{\textstyle\frac{3}{20}}g_t(\Lambda)^{-2}\ <\ 
{\textstyle\frac{2}{5}}g_1(\Lambda)^{-2}.\eee
We already had this same inequality \cite{cis} with
$z'=0$. For
$\Lambda=m_Z$ it means $\sin^2\theta_w<0.54$ and
remains harmless for higher $\Lambda$.

$\bullet$ To make contact with the noncommutative
Einstein-Hilbert action, we put $z=0$.  Now the
constraints on the gauge coupling are identical
to those from the Einstein-Hilbert action and force
upon us the big desert. In addition, the top mass is
constrained,
\bb g_t^2={\textstyle\frac{N}{3}}g_3^2.\eee
To compute the Higgs mass, we solve the system
(\ref{sys}), which is simple due to the
approximation of a dominating top mass: 
\bb \alpha =0,\qq \beta =0,\qq \gamma
=\,\frac{1}{2N}\, .\eee
From equation (\ref{lamL}) we have
\bb
\lambda(\Lambda)={{\textstyle\frac{3N-2}{24}}}\,g_3^2
={\textstyle\frac{7}{24}}\,g_3^2.\ee
This constraint is to be compared to the one from the
Einstein-Hilbert action (\ref{ehlam}), 
$\lambda(\Lambda)={\textstyle\frac{N}{9}}g_3^2$.
The two scalar selfcouplings would coincide precisely
if we had $N=6$ generations!

In terms of masses, we get in the soft case,
$\Lambda=0.96\cdot 10^{10}$ GeV:
\bb m_t&=&214\pm 0\pm 4\ {\rm GeV},\label{topsoft}\\
m_H&=& 227\pm 0\pm 4 \ {\rm GeV\qq \qq
soft\ YM},\ee
and in the stiff case, $z'\in\rho$(center):
\bb m_t&=&188\pm 14\pm 2\pm 0\ {\rm GeV},\label{topYMstiff}\\
m_H&=& 198\pm 8\pm 2 \pm 0\ {\rm GeV\qq \qq
stiff\ YM}.\label{higgsYMstiff}\ee
For comparison, we recall the values from the
Einstein-Hilbert action:
\bb m_H&=&190\pm 0\pm 1\pm 4\ {\rm GeV
\qq \qq soft\ EH},\label{topEHsoft}\\
 m_H&=& 182\pm 10\pm 2\pm 7 \ {\rm GeV\qq \qq
stiff\  EH}.\label{higgsEHstiff}\ee
The first error is from the
uncertainty in the noncommutative scale, 
 $\Lambda=(10^{13}-10^{17})$ GeV, the second from
the present experimental uncertainty in the gauge 
couplings,
$g_3=1.218\pm 0.026$, and the third is from the
uncertainty in the top mass, if needed as input,
$m_t=175\pm 6$ GeV.

We give up the soft noncommutative Yang-Mills model since its 
top mass (\ref{topsoft}) must be bigger than 197 GeV, Figure \ref{fig:The cutoff 
as function of the top mass}.
It is too large and we retain only the stiff model. 
The stiff values for the top mass from Yang-Mills (\ref{topYMstiff}) seem in contradiction 
with the asymptotic value 197 GeV from  Figure \ref{fig:The cutoff 
as function of the top mass}. They are not, the asymptotic value is sensitive to changes 
in the gauge couplings which  in the stiff case deviate slightly  from the experimental values. 
The error of $\pm 14$ GeV from the uncertainty in $\Lambda$ tells us how close we are 
to the rim. Concerning the Higgs mass, the two allowed intervals, (\ref{higgsEHstiff})
 from relativity and (\ref{higgsYMstiff}) from Yang-Mills have a non-empty intersection,
\bb
m_H = 188-201 \ {\rm GeV}.
\eee
This is the second pillar of the bridge we propose.
All values of top and Higgs masses (\ref{topsoft}-\ref{higgsEHstiff}) 
are compatible with perturbation and stability in the energy range, $m_Z<E<\Lambda$,
Figure \ref{fig:The allowed domains}.

\section{Conclusions}

There is an old hand waving argument combining
Heisenberg's uncertainty relation with the
Schwarzschild radius which implies that the
noncommutative scale $\Lambda$ must be smaller
than the Planck mass, $10^{19}$ GeV.
This is compatible with the numbers above and we
have the following picture. Grand unification
proposed the big desert, from the Higgs mass all the
way up to $\Lambda$ nothing new happens, no new
particle, not even a break down of perturbation theory
\cite{cab}. At $\Lambda$, a modest change of physics
happens. The standard $SU(3)\times SU(2)\times U(1)$
is unified into $SU(5)$. This adds a few more
Yang-Mills and Higgs bosons to our boring world.
These `lepto-quarks' cause proton decay and make our
world too exciting to be stable. Noncommutative
geometry also has the big desert, but on its other side a
revolution in form of a  truly noncommutative
space-time of which
 the cheap tensor product between differential
geometry and the internal space of the standard model
is only a low energy mirage. We expect that crossing
the scale
$\Lambda$ will induce `noncommutative' threshold
effects that are responsible for the small mismatch in 
the gauge couplings triangle in the
$E$-$g$ plane (Figure \ref{fig:The coupling constants}) 
and for the small mismatch in the
scalar selfcoupling and in the Yukawa coupling of the top. 
We also hope that the new
geometry beyond $\Lambda$ will solve our conceptual
problems with quantum field theory, in particular in
presence of gravity. But this is still troubled water. 

The dream to connect general relativity and
Yang-Mills theories is as old as Einstein. Elegant
attempts have been proposed, Kaluza-Klein theories,
Weyl's gravity, Poincar\'e gravity, Sakharov's induced
gravity... Noncommutative geometry proposes another
bridge, that stands so far thanks to a subtle
conspiration between the gauge couplings, the top
mass and the number of generations. But it will fall
soon if the Higgs mass does not cooperate.

\section{Appendix}

$\bullet$ {\bf Gauge couplings:} The $SU(2) \times U(1) \times SU(3)$ 
gauge couplings $ g_3,  g_1, g_2$ are normalized by the scalar product
\bb
<\rho (a, b, c), \rho (a, b, c)>_{z,z'} \ = \ \frac{1}{2} g_1^{-2} b\overline{b} + g_2^{-2} \t (a^{*}a) + g_3^{-2} \t (c^{*}c),
\eee
for $(a, b, c) \in su(2) \op u(1) \op su(3)$. \hfil \\
$\bullet$ {\bf Higgs field:} The kinetic term of the scalar field $\phi$ is normalized 
to $\frac{1}{2}$ in the Lagrangian which is written as
\bb
{\cal{L}} = \frac{1}{2} \partial_{\mu}\phi^*\partial_{\mu}\phi + 
\lambda (\phi^*\phi)^2 - \frac{\mu^2}{2}(\phi^*\phi) + ...
\eee
Moreover, if $v$ is its expectation value, the relations between the gauge 
couplings $g_{.}$, the selfcoupling $\lambda$ and the $W$, top, Higgs running
masses are defined by
\bb
v&=&2g_2^{-1}\ m_W, \nonumber \\
\lambda&=& \frac{g_2^2}{32}\ \frac{m_H^2}{m_W^2}, \nonumber \\
\mu&=& \frac{1}{\sqrt{2}}\ m_H, \nonumber \\
g_t&=&v^{-1} \ m_t.
\eee
All these relations depend on the energy. \hfil \\
$\bullet$ {\bf Renormalization procedure:} We adopt the mass 
independent ${\overline{MS}}$ renormalization scheme in the approximation 
where all fermions masses are neglected but the top quark mass $m_{t}$. As 
running parameter associated to the energy $E$, we choose 
$r=\log_{10}(\frac{E}{m_{Z}})$. For the renormalization flow, the 
one-loop evolution equations of the above variables are the 
following first order differential equations
\bb
C \ g_{1}'(r)&=&\frac{41}{6} g_{1}(r)^{3}, \label{g1evolution}\\
C \  g_{2}'(r)&=&-\frac{19}{6} g_{2}(r)^{3},  \label{g2evolution}\\
C \  g_{3}'(r)&=&-7 g_{3}(r)^{3}, \label{g3evolution}\\
C \  g_{t}'(r)&=&g_{t}(r)(-\frac{17}{12}g_{1}(r)^{2}-\frac{9}{4}g_{2}(r)^{2}-8g_{3}(r)^{2}+9g_{t}(r)^{2}) \\
C \ \lambda'(r)&=&\lambda(r)(-3g_{1}(r)^{2}-9g_{2}(r)^{2}+ 24 
g_{t}(r)^{2}+96 \lambda(r) ) \nonumber\\
&&\qq \qq + \frac{3}{32}g_{1}(r)^{4} +\frac{9}{32} 
g_{2}(r)^{4}-6g_{t}(r)^{4}+\frac{3}{16}g_{1}(r)^{2}g_{2}(r)^{2}, \\
C \ \mu'(r)&=&\mu(r)(-\frac{3}{4}g_{1}(r)^{2}-\frac{9}{4}g_{2}(r)^{2}+6g_{t}(r)^{2}+24 \lambda(r)),\label{muevolution}
\ee
with $C = \frac{16 \pi^{2}}{\ln(10)}$. \hfil \\
$\bullet$ {\bf Initial conditions:} At $r=0$, that is, at $m_{Z}=91.187$ GeV, we have \cite{table} 
\bb
g_{1}(0)&=&0.3575\pm 0.0001 , \label{initial g1} \\
g_{2}(0)&=&0.6507\pm 0.0007 , \label{initial g2}\\
g_{3}(0)&=&1.218\pm 0.0026 , \label{initial g3}\\
m_{t}(0)&=&175\pm 6\   {\rm GeV},\label{initial mt}\\
 m_{W}(0)&=&80.33\pm 0.15\   {\rm GeV}\label{initial mW}.
\ee
So we get for the central values
\bb
v(0) &=&246.903 \   {\rm GeV},\nonumber \\
g_{t}(0) &=&0.0040. \nonumber 
\eee

The last equation (\ref{muevolution}) decouples from the others: note that 
$g_{1},g_{2},g_{3},g_{t},\lambda$ have no dimension while $\mu$ is a mass. 
At this point, it is important to quote that in our renormalization 
scheme, quadratic divergences do not appear, only the 
logarithmic ones are retained. In order to avoid renormalization scheme
ambiguities in the evolution of $\mu^2$, we neglect
the threshold effects of the top and Higgs masses and
we identify their pole masses
$m_p=m(m_p)$ with their running
masses at the $Z$ mass $m(m_Z)$. 
For these reasons, we will not use (\ref{muevolution}).
\hfil \\
$\bullet$ {\bf Figures:} All quantities not explicitely mentioned in a 
figure are put to their experimental values.

\begin{figure}[hbt]
\hspace{2cm}
\def\epsfsize#1#2{0.6#1}
\epsfbox{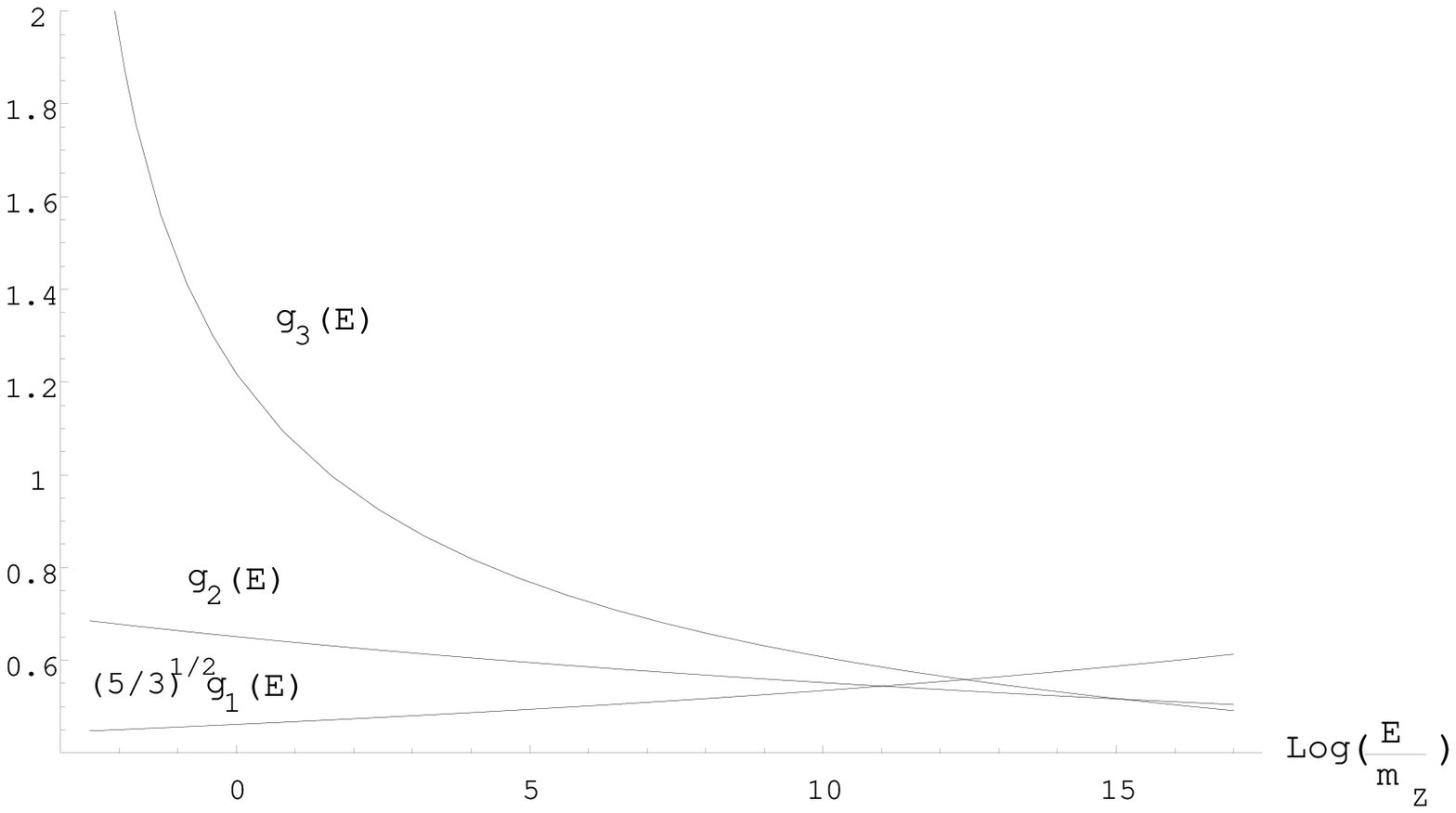}
\caption{The evolution of the three coupling constants}
\label{fig:The coupling constants}
\end{figure}

\begin{figure}[hbt]
\hspace{2cm}
\def\epsfsize#1#2{0.6#1}
\epsfbox{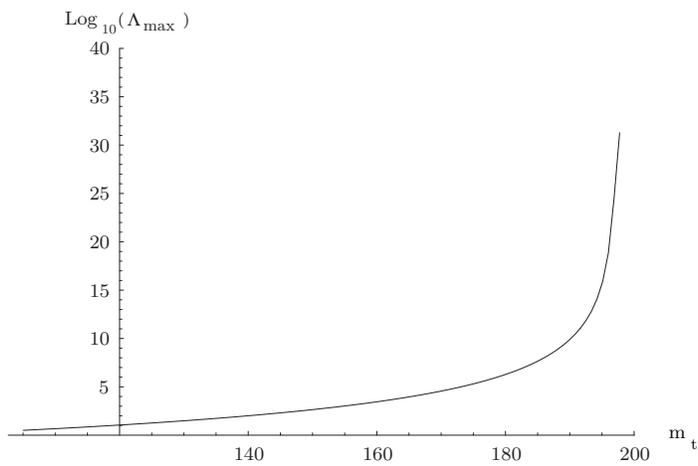}
\caption{The cutoff $\Lambda_{\rm max}$ as function of the top mass}
\label{fig:The cutoff as function of the top mass}
\end{figure}

\begin{figure}[hbt]
\hspace{2cm}
\def\epsfsize#1#2{0.6#1}
\epsfbox{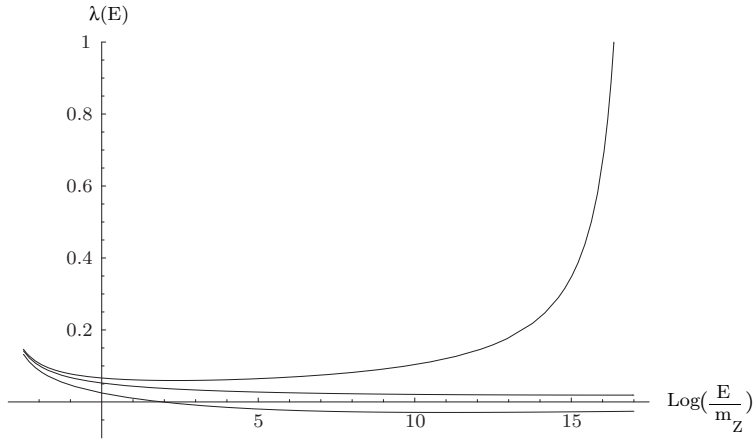}
\caption{The Higgs selfcoupling for $m_H(m_Z) = 120$ (lower graph), 
$160$ and $180$ Gev (upper graph) for $m_t(m_Z) = 175$ GeV}
\label{fig:The Higgs selfcoupling}
\end{figure}

\begin{figure}[hbt]
\hspace{2cm}
\def\epsfsize#1#2{0.6#1}
\epsfbox{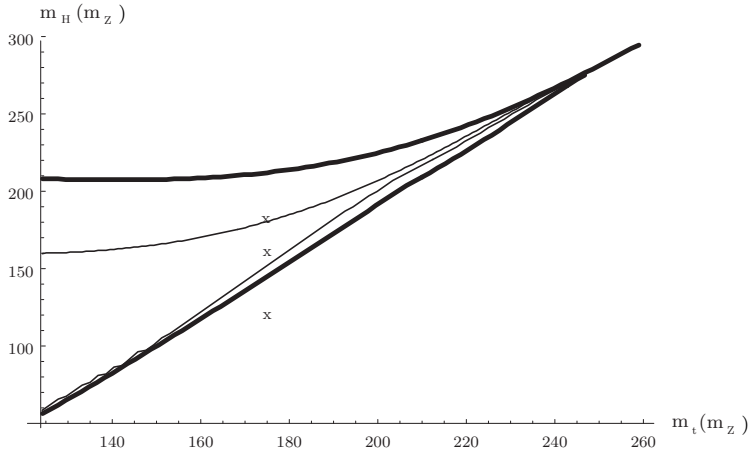}
\caption{Two allowed domains for the Higgs mass with $\Lambda= 10^{10}$ GeV (thick lines) and 
$\Lambda=10^{19}$ GeV (thin lines): slices between the upper curves ($\lambda<1$) and 
the lower curves ($\lambda>0$), 
with points drawn at $m_H(m_Z) = 120, 160, 
180$ GeV, for $m_t(m_Z) = 175$ GeV}
\label{fig:The allowed domains}
\end{figure}

In Figure \ref{fig:The coupling constants}, the intersections of the three coupling 
constants determine a triangle. Figure \ref{fig:The top 
coupling} shows the evolution of $g_{t}$. Clearly, the same evolution for 
the Higgs selfcoupling $\lambda$ strongly depends in Figure \ref{fig:The Higgs selfcoupling} 
of the Higgs initial value mass. This is due to the fact that the allowed domain for the Higgs 
mass in term of the top mass (Figure \ref{fig:The allowed domains}) is the slice between the 
top curve which describes the perturbative (or triviality) condition 
($\lambda <1$) and the down curve which 
describes the instability condition ($\lambda >0$) for energies between the $Z$ mass 
and the Planck mass. Naturally, this slice depends on the choice of $\Lambda$. 
The three points in Figure \ref{fig:The allowed domains} 
are the initial conditions of the curves of Figure \ref{fig:The Higgs selfcoupling}. 
The upper point corresponds to the upper curve which is non perturbative and 
the lower point corresponds to the lower curve which is unstable.
Since the experimental top mass 
(\ref{initial mt}) localizes the Higgs mass in a very narrow region, 
this figure is particularly significant if we believe in  
perturbation and stability throughout the big desert.
It is important to note here that these 
two assumptions traditionally imposed by hand in a classical Yang-Mills-Higgs theory, 
are automatically satisfied in both noncommutative dreis\"atze: once 
the initial conditions (\ref{initial g1}-\ref{initial mW}) are admitted,  
the selfcoupling $\lambda$ always stays in the stable and perturbative regime for 
energies between $m_Z$ and $\Lambda$. Note for instance that, in the soft case, 
$\Lambda=0.96\cdot 10^{10}$ GeV is small compared to the Planck mass.
Figure \ref{fig:The cutoff as function of the top mass} shows again that the noncommutative cutoff 
$\Lambda_{\rm max}$ is very sensitive to a top mass around $197$ GeV.

\vfil

\begin{figure}[hbt]
\hspace{2cm}
\def\epsfsize#1#2{0.6#1}
\epsfbox{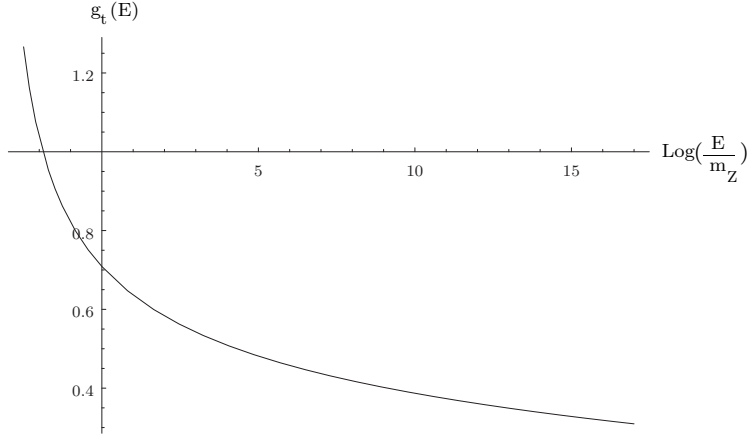}
\caption{The top coupling}
\label{fig:The top coupling}
\end{figure}

 \end{document}